\documentclass{elsart}
\pdfoutput=1
\usepackage{graphicx}
\usepackage{amssymb}
\usepackage{amsmath}
\usepackage{tikz}
\usetikzlibrary{shapes,arrows}
\begin{document}

\tikzstyle{decision} = [diamond, draw, fill=red!20, text width=7em, text badly centered, node distance=4cm, inner sep=0pt]
\tikzstyle{block} = [rectangle, draw, fill=blue!20, text width=10em, text centered, rounded corners, minimum height=4em]
\tikzstyle{line} = [draw, very thick, color=black!50, -latex']
\tikzstyle{cloud} = [draw, ellipse,fill=yellow!20, node distance=2.5cm, minimum height=2em]

\begin{frontmatter}
\title{Modelling of the motion of a Mecanum-wheeled vehicle}
\author{E. Matsinos{$^*$}}
\address{Institute of Mechatronic Systems, Zurich University of Applied Sciences (ZHAW), Technikumstrasse 5, P.O. Box, CH-8401 Winterthur, Switzerland}

\begin{abstract}The present document aims at developing the formalism needed in order to describe the two-dimensional motion of a vehicle with Mecanum wheels, including the effects of resistive friction. The description of 
recently-acquired experimental data, on the basis of this model, is found satisfactory.\\
\noindent {\it PACS:} 
\end{abstract}
{$^*$}{E-mail: evangelos.matsinos@zhaw.ch, evangelos.matsinos@sunrise.ch; Tel.: +41 58 9347882; Fax: +41 58 9357780}
\end{frontmatter}

\section{\label{sec:Introduction}Introduction}

Despite their use in several applications, few efforts have been taken toward the theoretical description of the motion of Mecanum-wheeled systems. One recent article, dealing with the modelling of such systems \cite{tv}, has 
not been edited to the extent of eliminating errors and misprints in the formulae it contains. A follow-up paper by the same authors \cite{vt} does constitute an improvement, yet it is not detailed enough in relation to the 
results obtained in that study. At the end their paper \cite{vt}, the authors repeat a statement which appeared in an earlier version of the work \cite{vb}, namely that `More tests will be run on the single wheel tester to 
eliminate some of the errors seen in the current results'. However, the reader of these papers cannot (at least, in a direct manner) obtain a clear idea of the magnitude and seriousness of the `errors seen in the current 
results' in terms of the description of the experimental data with the proposed model.

The present paper serves two purposes: a) to provide a firm basis for the description of the motion of a Mecanum-wheeled vehicle and b) to establish an easy-to-implement solution, which could be used in the study of such systems.

The mathematical basis for the description of mobile robots was developed by Muir and Neuman in the mid 1980s \cite{mn}; in that work, the authors had modelled the motion for six prototype wheeled mobile robots in terms 
of their so-called wheel equations. Despite its formality, however, the Muir-Neuman paper does not provide treatment of the effects of the resistive friction.

The outline of the present paper is as follows. Section \ref{sec:Method} will provide the details of the formalism which has been developed in order to model the motion of a Mecanum-wheeled vehicle on a horizontal plane; the motion 
will be determined in the general case, i.e., including resistive friction. Section \ref{sec:ER} will confront the model of Section \ref{sec:Method} with recently-acquired experimental data; the first part of the section will provide 
a short description of the experiments performed, whereas the second part will give the results of the analysis of the experimental data. Finally, the conclusions of the present work will be given in Section \ref{sec:Conclusions}. 
It should be borne in mind, right from the beginning, that, as this study relies exclusively on very limited and specific data, it should not be regarded as complete; the proposed model must be confronted with experimental data, 
pertaining to arbitrary, not only translational motion.

\section{\label{sec:Method}Method}

\subsection{\label{sec:General}General}

\begin{figure}
\begin{center}
\includegraphics [width=7.5cm,angle=270] {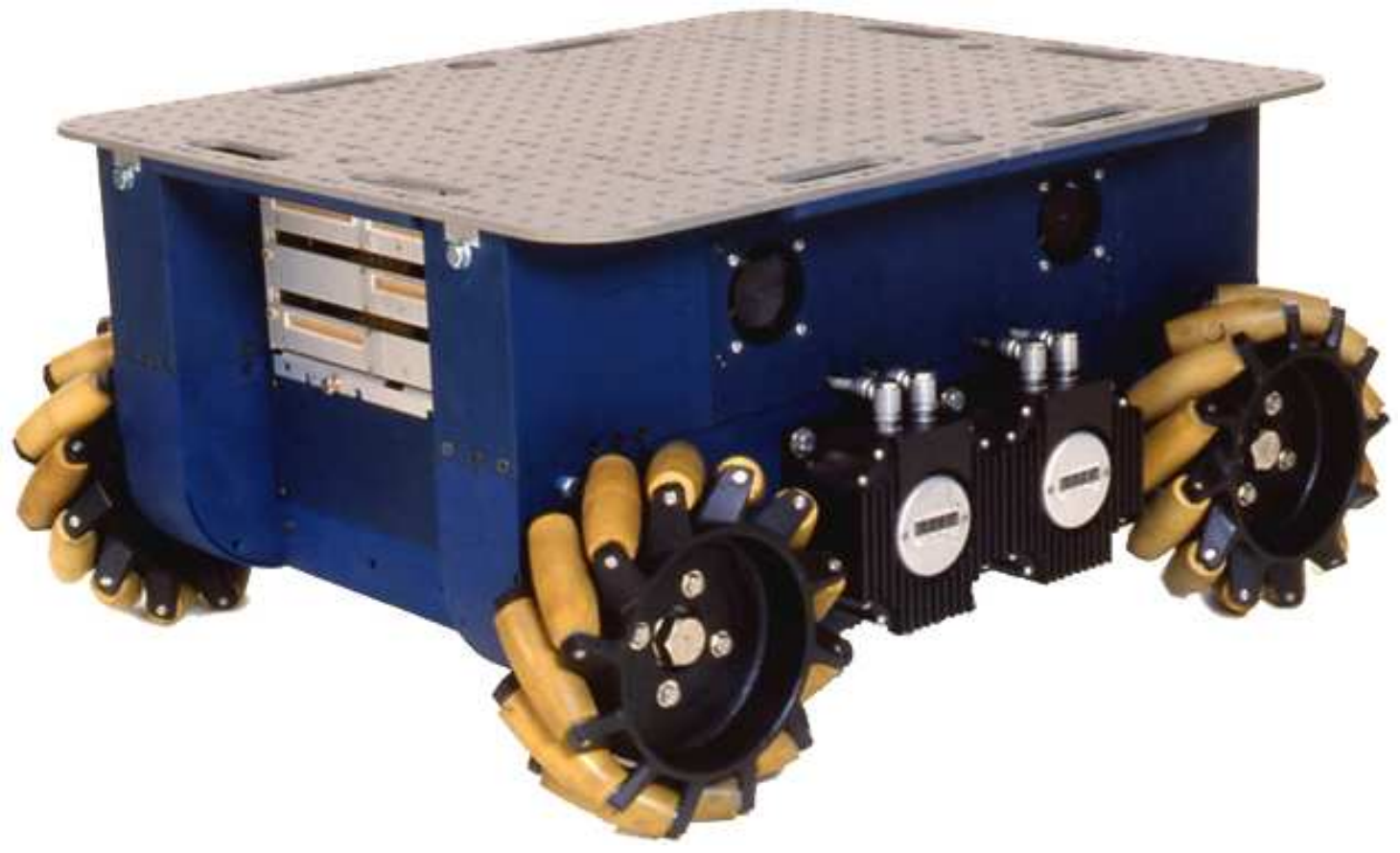}
%\vspace{-6cm}
\caption{\label{fig:Uranus}The mobile robot URANUS \cite{ur} utilising Mecanum wheels for omni-directional motion.}
\end{center}
\end{figure}

The Mecanum wheel is frequently called `the Ilon wheel' after its inventor, Bengt Ilon, who came up with the original design in 1973 while being employed at the Swedish company Mecanum AB. It is a conventional wheel with a number 
of rollers attached to its circumference. The axis of rotation of the rollers is inclined with respect to the rotational `plane' of the wheel, in a plane parallel to the axis of rotation (of the wheel). Omni-directional motion of 
Mecanum-wheeled vehicles is achieved by appropriately controlling the angular velocities of each wheel, as well as the direction of rotation (we will assume herein that each wheel is powered by a dedicated motor); in the general 
case, the motion is a combination of rotation and translation. If all wheels are turned in the same direction, at equal angular velocity, forward/backward motion of the vehicle is achieved. By rotating the wheels on the same side 
against each other, a sideways motion of the vehicle is achieved. There are velocity/rotation combinations resulting only in rotation of the vehicle. The mobile robot URANUS \cite{ur} (Fig.~\ref{fig:Uranus}), the first 
flexible mobile robot designed and constructed in order to provide a general-purpose mobile base to support research in robot navigation, used Mecanum wheels for omni-directional motion. An overview of Mecanum-wheeled systems, albeit 
avoiding the mathematical intricacies, including a discussion on their application, advantages, and disadvantages, may be found in Ref.~\cite{ad}.

\subsection{\label{sec:Formalism1}Modelling of the motion of a vehicle with four Mecanum wheels: frictionless case}

Generally speaking, it does not sound logical to talk about motion of vehicles in the case of vanishing friction; it is through friction that motion is enabled. What is actually implied when using the term `frictionless' in the 
description of a motion is that the tractive forces are large enough to enable motion, whereas the resistive ones are conveniently small so that they may safely be ignored. In this section, the formalism needed in order to describe 
such an ideal situation will be developed.

\begin{figure}
\begin{center}
\includegraphics [width=14cm] {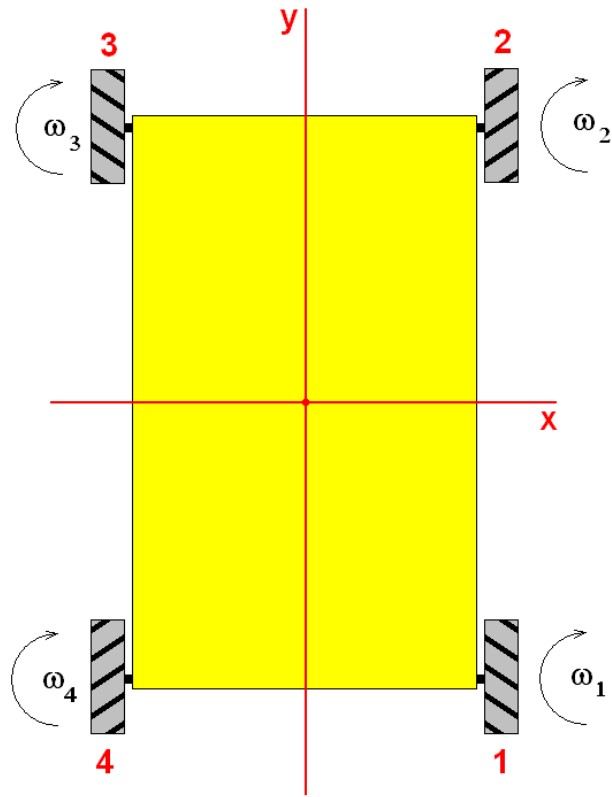}
%\vspace{-6cm}
\caption{\label{fig:Bogie1}Top view of a vehicle with four Mecanum wheels, along with its attached coordinate system ($x$,$y$).}
\end{center}
\end{figure}

Figure \ref{fig:Bogie1} provides a top view of a (rectangular) vehicle featuring four Mecanum wheels, along with its attached coordinate system ($x$,$y$), the origin of which is assumed to be the geometrical centre of the rectangle; 
the wheels are identified by the numbers $1 \dots 4$, starting from the right-bottom corner (i.e., from the right-rear wheel of the vehicle) and proceeding in the counter-clockwise direction. The angular velocities $\omega_{1 \dots 4}$ 
are defined positive for translational motion in the forward direction (increasing $y$).

\begin{figure}
\begin{center}
\includegraphics [width=14cm] {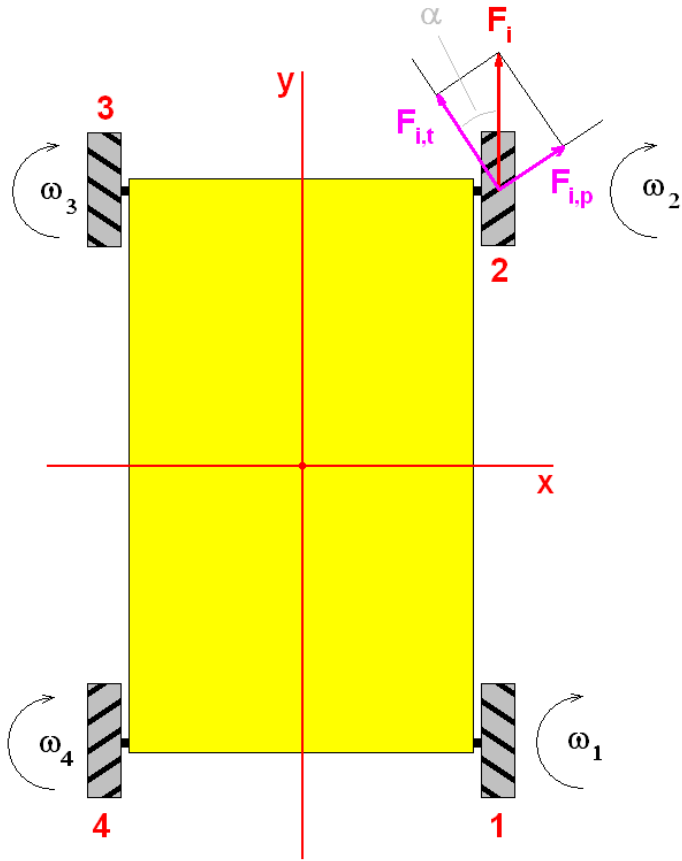}
%\vspace{-6cm}
\caption{\label{fig:Bogie2}The driving (motor) force $F_i$ acting on one of the wheels of the vehicle (chosen to be wheel $2$), along with its decomposition into one force ($F_{i,p}$) parallel to the rotational axis of the roller, which is in contact 
with the ground at that moment, and one in the transverse direction ($F_{i,t}$). The angle between the transverse component $F_{i,t}$ and the rotational `plane' of the wheel is denoted as $\alpha$. The rollers shown are assumed to be those corresponding to the 
lower part of the wheel, part of which is in contact with the ground.}
\end{center}
\end{figure}

The driving (motor) force (thrust) $\vec{F}_i$ acting on wheel $i$ of the vehicle (chosen to be wheel $2$), along with its decomposition into one force ($\vec{F}_{i,p}$) parallel to the rotational axis of the roller (which is in 
contact with the ground at that moment) and one in the transverse direction ($\vec{F}_{i,t}$), are shown in Fig. \ref{fig:Bogie2}. The angle between the transverse direction and the rotational plane of the wheel is denoted as 
$\alpha \in [0,\pi)$. (The quantity $\sin \alpha$ is also known as the `efficiency of the wheel'.) Since the rollers rotate freely around their axle, there is no traction along the transverse direction; therefore, the force 
$\vec{F}_{i,t}$ can safely be ignored when studying the motion of the vehicle. The relation between $F_{i,p}$ and $F_i$ (indicating the corresponding moduli of the two vectors) reads as: $F_{i,p}=F_i \sin \alpha$.

\begin{figure}
\begin{center}
\includegraphics [width=14cm] {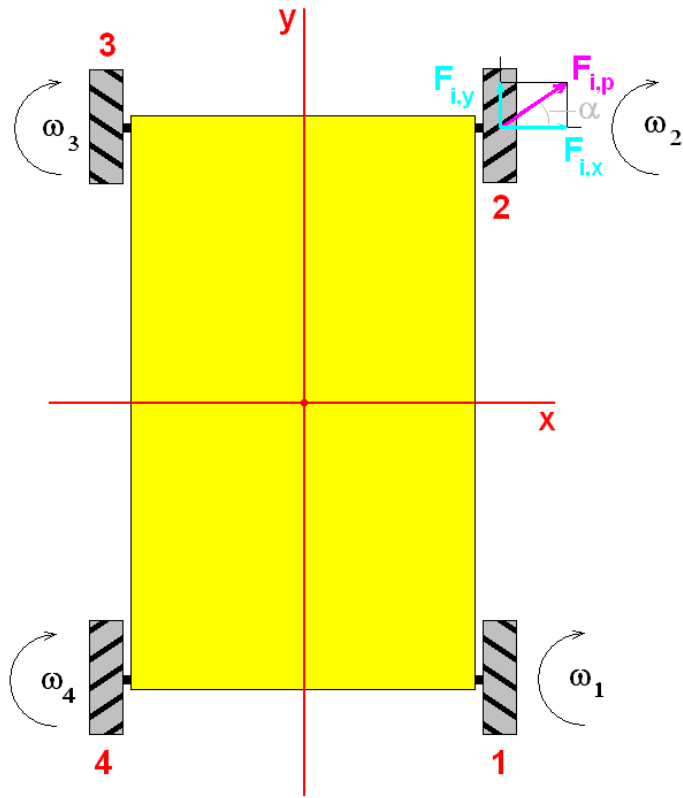}
%\vspace{-6cm}
\caption{\label{fig:Bogie3}Decomposition of the force $F_{i,p}$ of Fig.~\ref{fig:Bogie2} into two forces $F_{i,x}$ and $F_{i,y}$, parallel to the axes of the attached coordinate system. The rollers shown are assumed to be those corresponding to the 
lower part of the wheel, part of which is in contact with the ground.}
\end{center}
\end{figure}

Finally, the only relevant force, $\vec{F}_{i,p}$, may be decomposed into forces along the axes of the attached coordinate system (see Fig.~\ref{fig:Bogie3}). The geometry dictates that 
$F_{i,x}=F_{i,p} \cos \alpha=F_i \sin \alpha \, \cos \alpha$ and $F_{i,y}=F_{i,p} \sin \alpha=F_i \sin^2 \alpha$.

\begin{figure}
\begin{center}
\includegraphics [width=14cm] {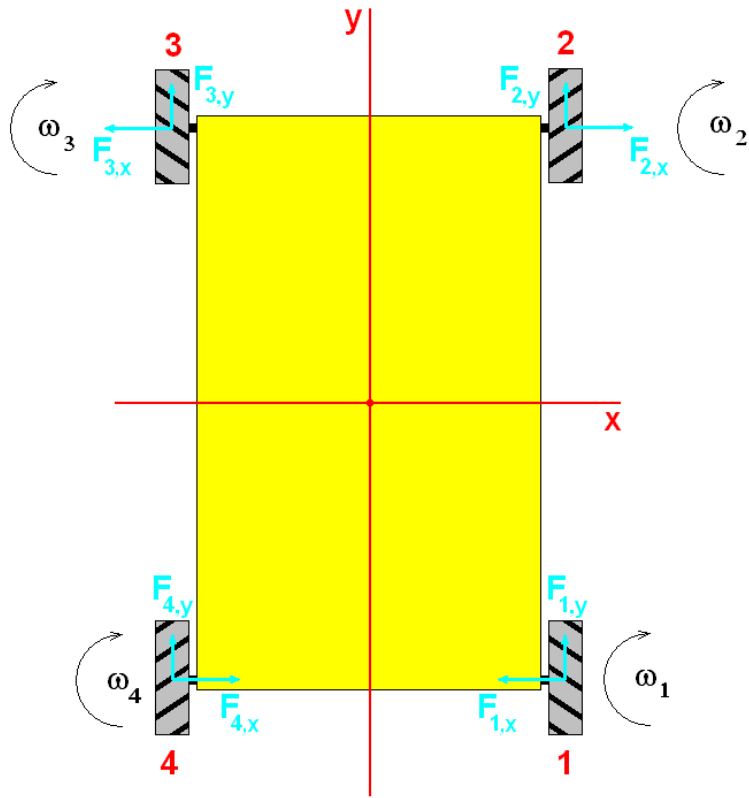}
%\vspace{-6cm}
\caption{\label{fig:Bogie4}Cancellation of the lateral forces in the special case $\omega_1=\omega_2=\omega_3=\omega_4$. The rollers shown are assumed to be those corresponding to the lower part of the wheel, part of which is in contact with the ground.}
\end{center}
\end{figure}

To obtain the expression for the total force acting on the vehicle in the $x$ and $y$ directions, one must consider the different orientations of the axles of the rollers on each wheel; in the configuration which is `casually' 
used, these orientations are identical for diagonal wheels and opposite (mirror images) for the two front/back wheels. From now on, the sign of $F_i$ will be omitted (i.e., the tractive force will always be assumed positive) and 
the direction of applied force will instead be taken into account in the rotational direction of each wheel~\footnote{The projection of the angular velocity $\overrightarrow{\omega_i}$ along the $x$ direction (simply denoted as 
$\omega_i$) is either positive or negative.}. Assuming that the quantity $\alpha$ represents the orientation of the roller axes of wheels $2$ and $4$, one may easily deduce that the force acting on the vehicle in the $x$ direction is:
\begin{equation} \label{eq:FX}
F_x=\sum_{i=1}^{4} F_{i,x}=\sin \alpha \, \cos \alpha \sum_{i=1}^{4} (-1)^i sgn(\omega_i) F_i \, \, \, ;
\end{equation}
the force in the $y$ direction reads as:
\begin{equation} \label{eq:FY}
F_y=\sum_{i=1}^{4} F_{i,y}=\sin^2 \alpha \sum_{i=1}^{4} sgn(\omega_i) F_i \, \, \, .
\end{equation}
It must be mentioned that Eqs.~($9$, $10$) of Ref.~\cite{tv} contain one additional quantity ($K_i$), introduced as `the wheel constant, dependent on the number of rollers per wheel and [on] how tight the rollers are [fixed] on 
the wheel's hub'. Given that the goal herein is to develop the formalism for similar wheels (material and number of rollers, as well as inclination $\alpha$), the $K_i$'s can be assumed constant and, as such, be absorbed in 
a redefinition of the tractive forces $F_i$. Finally, the total force acting on the vehicle is given by
\begin{equation} \label{eq:FT}
\vec{F}=F_x \hat{u}_x + F_y \hat{u}_y \, \, \, ,
\end{equation}
where $\hat{u}_x$ and $\hat{u}_y$ denote the unit vectors along the $x$ and $y$ directions, respectively. In the special case that $\omega_1=\omega_2=\omega_3=\omega_4$, the lateral forces cancel out (i.e., the force in the $x$ 
direction vanishes, see Fig.~\ref{fig:Bogie4}).

The motion of a rigid body in two dimensions is a superposition of a) a translational motion of its centre of mass (CM) and b) a rotation around the axis passing through the body's CM, normal to the plane on which the motion is confined.

To determine the translational motion of the vehicle in the laboratory, one needs to obtain the components of the force in that reference frame from those of the CM coordinate system. If the latter is rotated by an angle $\phi$ with respect to 
the former (a parallel translation leaves the orientation angle of a vector invariant), the transformation reads as
\begin{equation} \label{eq:FLAB}
\left( \begin{array}{c} F_X \\ F_Y \end{array} \right) = \left( \begin{array}{ccc} \cos \phi & \, \, & -\sin \phi \\ \sin \phi & \, \, & \cos \phi \end{array} \right) \, \left( \begin{array}{c} F_x \\ F_y \end{array} \right) \, \, \, .
\end{equation}
The total force $\vec{F}_L$ in the laboratory coordinate system is given by
\begin{equation} \label{eq:FTLAB}
\vec{F}_L=F_X \hat{u}_X + F_Y \hat{u}_Y \, \, \, ,
\end{equation}
where $\hat{u}_X$ and $\hat{u}_Y$ denote the unit vectors along the $X$ and $Y$ directions.

\begin{figure}
\begin{center}
\includegraphics [width=14cm] {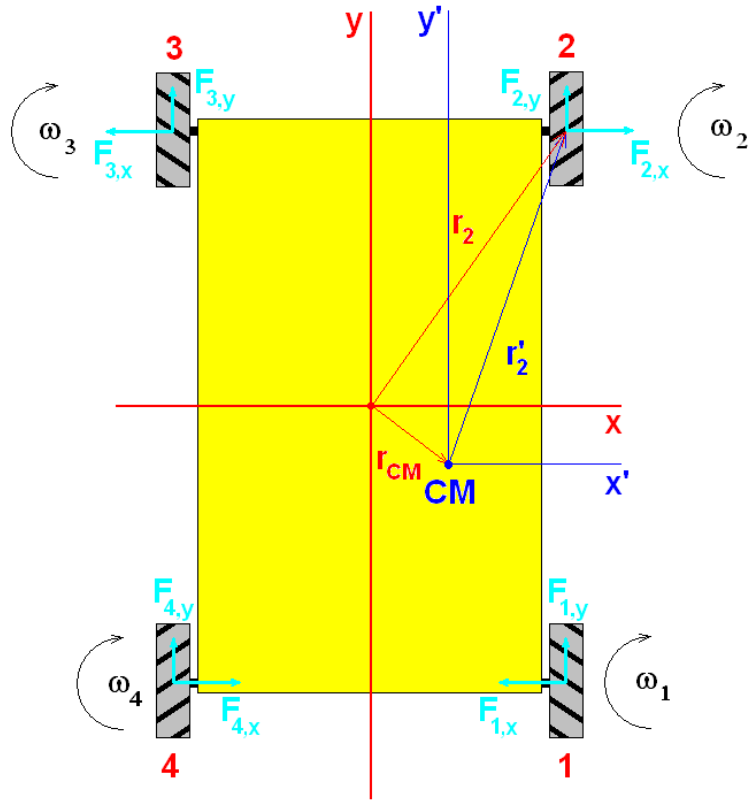}
%\vspace{-6cm}
\caption{\label{fig:Bogie5}The attached coordinate system ($x$,$y$), along with the one having its origin at the centre of mass (CM) of the vehicle ($x^\prime$,$y^\prime$). The rollers shown are assumed to be those corresponding to 
the lower part of the wheel, part of which is in contact with the ground.}
\end{center}
\end{figure}

The translational motion of the vehicle in the laboratory coordinate system is obtained on the basis of $\vec{F}_L$ of Eq.~(\ref{eq:FTLAB}) and of the mass $m$ (also containing the equivalent mass of loads) of the vehicle, whereas 
the rotation will involve the torque $\vec{T}$ acting on the vehicle.
\begin{align} \label{eq:TQ}
\vec{T} = & \sum_{i=1}^{4} \vec{r_i^\prime} \times \vec{F_i} = \sum_{i=1}^{4} (\vec{r_i}-\overrightarrow{r_{CM}}) \times \vec{F_i} = \sum_{i=1}^{4} \vec{r_i} \times \vec{F_i} - \overrightarrow{r_{CM}} \times \sum_{i=1}^{4} \vec{F_i} = \nonumber \\
& \sum_{i=1}^{4} \vec{r_i} \times \vec{F_i} - \overrightarrow{r_{CM}} \times \vec{F} \, \, \, ,
\end{align}
where $\overrightarrow{r_{CM}}$ identifies the vehicle's CM in the attached coordinate system ($x$,$y$), $\vec{r_i}$ the position vector of wheel $i$ in ($x$,$y$), and $\vec{r_i^\prime}$ the position vector of wheel $i$ in the coordinate 
system ($x^\prime$,$y^\prime$) (see Fig.~\ref{fig:Bogie5}). The torque $\vec{T}$ will generate a rotation around the $z^\prime$-axis of the coordinate system, passing through the CM of the vehicle; the rotation obeys the equation
\begin{equation} \label{eq:N2}
\vec{T} = I \ddot{\phi} \hat{u}_{z^\prime} \, \, \, ,
\end{equation}
where $I$ is the moment of inertia of the vehicle (around $z^\prime$), $\ddot{\phi}$ represents the angular acceleration, and $\hat{u}_{z^\prime}$ is the unit vector along the $z^\prime$ direction 
($\hat{u}_{z^\prime} = \hat{u}_{x^\prime} \times \hat{u}_{y^\prime}$).

\subsection{\label{sec:Formalism2}Determination of the position of the vehicle}

\begin{figure}
\begin{center}
\includegraphics [width=14cm] {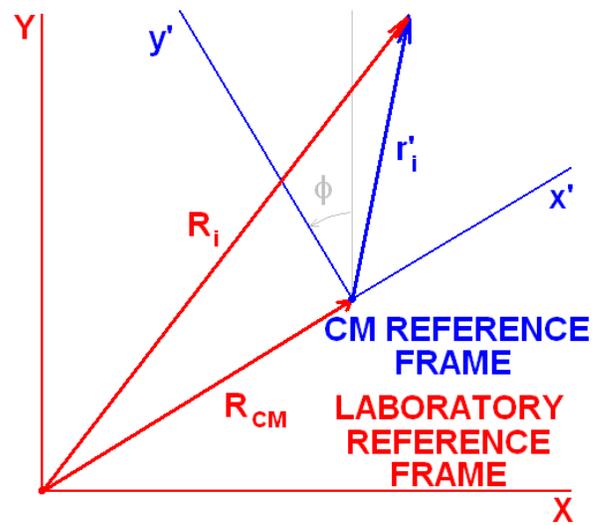}
%\vspace{-6cm}
\caption{\label{fig:Bogie6}The relation between the laboratory coordinate system ($X$,$Y$) and the centre-of-mass (CM) coordinate system ($x^\prime$,$y^\prime$) of the vehicle.}% The position (and orientation) of the vehicle is known at all times $t$, if the 
%functions $\overrightarrow{R_{CM}} (t)$ and $\phi (t)$ have been obtained.}
\end{center}
\end{figure}

The problem of the determination of the position of the vehicle as a function of time $t$ will be split up into two parts. a) Evaluation of the position of the vehicle's CM as a function of time (i.e., determination of the function 
$\overrightarrow{R_{CM}} (t)$, where $\overrightarrow{R_{CM}}$ denotes the position vector of the CM in the laboratory coordinate system). b) Evaluation of the orientation angle of the vehicle (i.e., of the amount of the rotation around the 
$z^\prime$-axis) as a function of time (i.e., determination of the function $\phi (t)$). A schematic form of the relation between these two coordinate systems is shown in Fig.~\ref{fig:Bogie6}.

Let us assume that the position of the CM of the vehicle at $t=t_1$ is known (i.e., that $\overrightarrow{R_{CM}} (t_1)$ is given), as is the velocity $\overrightarrow{V_{CM}} (t_1)$. An estimation of $\overrightarrow{R_{CM}}(t_1+\Delta t)$ and 
$\overrightarrow{V_{CM}}(t_1+\Delta t)$ will first be obtained. The time interval $\Delta t$ is chosen small enough to ensure that the acceleration between $t_1$ and $t_1 + \Delta t$ be safely considered constant. 
If $\overrightarrow{A_{CM}}=\vec{F}_L/m$, with $\vec{F}_L$ taken from Eq.~(\ref{eq:FTLAB}), one obtains
\begin{equation} \label{eq:VCM}
\overrightarrow{V_{CM}}(t_1 + \Delta t) = \overrightarrow{V_{CM}}(t_1) + \overrightarrow{A_{CM}} \Delta t
\end{equation}
and
\begin{equation} \label{eq:RCM}
\overrightarrow{R_{CM}}(t_1 + \Delta t) = \overrightarrow{R_{CM}}(t_1) + \overrightarrow{V_{CM}}(t_1) \Delta t + \frac{1}{2} \overrightarrow{A_{CM}} (\Delta t)^2 \, \, \, .
\end{equation}

The corresponding expressions for the angular velocity and angle may easily be obtained by using the standard substitutions: $\vec{F} \rightarrow \vec{T}$, $\vec{v} \rightarrow \vec{\omega}$, $\vec{r} \rightarrow \vec{\phi}$, and $m \rightarrow I$. Therefore,
\begin{equation} \label{eq:Omega}
\vec{\omega}(t_1 + \Delta t) = \vec{\omega} (t_1) + \frac{\vec{T}}{I} \Delta t
\end{equation}
and
\begin{equation} \label{eq:Phi}
\vec{\phi}(t_1 + \Delta t) = \vec{\phi}(t_1) + \vec{\omega}(t_1) \Delta t + \frac{1}{2} \frac{\vec{T}}{I} (\Delta t)^2 \, \, \, .
\end{equation}

\subsection{\label{sec:Formalism3}Modelling of the motion of a vehicle with four Mecanum wheels: introduction of resistive friction}

Resistive friction is generally categorised as static and dynamic.
\begin{itemize}
\item Static resistive friction results in the force between two objects in contact when they are not in motion relative to each other. The static friction prevents the \emph{start-up} of motion, as long as the tractive force $F_m$ 
remains smaller than a fraction of the (modulus of the) normal force $N$ exerted between the two objects, i.e., as long as $F_m \leq F_{\sigma} \equiv \mu_{\sigma} N$; in this relation, $\mu_{\sigma}$ is identified as the coefficient of static friction.
\item Dynamic resistive friction occurs when two objects in contact are moving relative to (i.e., rubbing against) each other; the coefficient of the dynamic friction will be denoted as $\mu_{\kappa}$.
\end{itemize}
The coefficient $\mu_{\kappa}$ is usually considered to be smaller than $\mu_{\sigma}$ (for the same materials), though this condition does not qualify as a necessary theoretical constraint; throughout the present work, it will be 
assumed that $\mu_{\sigma} \equiv \mu_{\kappa}$.

One must distinguish between two types of resistive friction, namely sliding and rolling; the former pertains to translational motion, the latter to rotational. In both cases, no distinction will be made between static and dynamic effects. The coefficient 
of sliding friction will be denoted as $\mu_s$, the one of rolling friction as $\mu_r$.

In addition to the aforementioned types of resistive forces, the viscous resistive friction (introduced in analogy to the force caused by the viscosity of lubricants) will be taken into account; it depends linearly on the relative velocity $\vec{v}$ 
between the two objects in contact: $\vec{F}_v = - \mu_v N \vec{v}$, where $\mu_v$ is the viscous-friction coefficient. No distinction will be made between sliding and rolling viscous resistive friction. As a matter of fact, the introduction of the 
viscous resistive friction in the present work is mathematically equivalent to assuming that the sliding- and rolling-friction coefficients are not constants, but depend (linearly, and in the same manner) on the velocity $\vec{v}$.

Let us denote the velocity of wheel $i$ in the laboratory coordinate system at a time instant $t$ as $\vec{V}_i$. Evidently,
\begin{align} \label{eq:VI}
\vec{V}_i = & \overrightarrow{V_{CM}} + {\bf R}(\phi) \big[ \vec{\omega} \times \vec{r_i^\prime} \big] = \nonumber \\
& \overrightarrow{V_{CM}} - \dot{\phi} \big[ (y_i^\prime \cos \phi + x_i^\prime \sin \phi) \hat{u}_X + (y_i^\prime \sin \phi - x_i^\prime \cos \phi) \hat{u}_Y \big] \, \, \, ,
\end{align}
where ${\bf R}(\phi)$ denotes the $2 \times 2$ rotation matrix appearing on the right-hand side of Eq.~(\ref{eq:FLAB}).

Each vector $\vec{V}_i$ may be decomposed as: $\vec{V}_i = V_{i,X} \hat{u}_X + V_{i,Y} \hat{u}_Y$. To determine the direction of the resistive force on wheel $i$, one must take three factors into account: a) the orientation of the vehicle in the 
laboratory coordinate system (angle $\phi$), b) the orientation of the axle of the rollers with respect to the rotational plane of the wheel, and c) the direction of rotation imposed on the wheel. For wheels $1$ and $3$, the orientation angle (of 
the axes of the rollers) in the laboratory coordinate system (i.e., the angle with the $X$-axis) is equal to $\pi - \alpha + \phi$; for wheels $2$ and $4$, it is equal to $\alpha + \phi$. Therefore, the directional vector for wheels $2$ and $4$ is 
$\hat{u}_{i,p}=\cos(\alpha+\phi) \hat{u}_X+\sin(\alpha+\phi) \hat{u}_Y$; for wheels $1$ and $3$, $\hat{u}_{i,p}=\cos(\pi-\alpha+\phi) \hat{u}_X+\sin(\pi-\alpha+\phi) \hat{u}_Y$. The projection of the vector $\vec{V}_i$ along the appropriate direction 
will determine the sign of the components of the resistive force (we will deal with its modulus $\mid \vec{f_i} \mid$ shortly), namely of $f_{i,X}$ and $f_{i,Y}$. For wheels $2$ and $4$, the projection of the vector $\vec{V}_i$ along the $\hat{u}_{i,p}$ 
direction is equal to $V_{i,p} = V_{i,X} \cos (\alpha + \phi) + V_{i,Y} \sin (\alpha + \phi)$; for wheels $1$ and $3$, $V_{i,p} = V_{i,X} \cos (\pi - \alpha + \phi) + V_{i,Y} \sin (\pi - \alpha + \phi)$. Therefore, if $\alpha \in (0,\frac{\pi}{2})$, the 
two components of the resistive force have the same sign (i.e., either both positive or both negative) for wheels $2$ and $4$; for wheels $1$ and $3$, they are of opposite sign.

For the evaluation of the modulus of the resistive force, one must first determine the corresponding normal force $N_i$ acting on wheel $i$. For planar motion of the vehicle, 
\begin{equation} \label{eq:FN}
N_i = N \frac{L_x - \mid x_i^\prime \mid}{L_x} \, \frac{L_y - \mid y_i^\prime \mid}{L_y} \, \, \, ,
\end{equation}
where $L_x$ and $L_y$ are the distances between the wheels in the $x$ and $y$ directions (`track' and `wheelbase' of the vehicle, respectively); the coordinates of wheel $i$ ($x_i^\prime$ and $y_i^\prime$) correspond to the CM coordinate 
system (see Fig.~\ref{fig:Bogie5}). If $\overrightarrow{r_{CM}}=\vec{0}$, all forces $N_i$ have the same magnitude (i.e., one quarter of the total load $N$). To obtain $f_i$ from $N_i$, one must multiply by the appropriate coefficient of friction, 
i.e., either $\mu_s$ (if the wheel $i$ is not moving in the direction of the traction) or $\mu_r$ (if it is); if the wheel is in motion, the viscous-friction force must also be included. Finally, the forces $F_x$ and $F_y$ of Eqs.~(\ref{eq:FX}, 
\ref{eq:FY}) will be redefined after taking into account the contributions of the resistive forces.

At this point, one remark is due. Rewriting Eq.~(\ref{eq:FX}) and absorbing each $sgn(\omega_i)$ factor in the corresponding $F_i$, one obtains: $F_x=(-F_1+F_2-F_3+F_4) \sin \alpha \cos \alpha$; Equation (\ref{eq:FY}) leads to: 
$F_y=(F_1+F_2+F_3+F_4) \sin^2 \alpha$. Assuming no offset of the CM of the vehicle with respect to the attached coordinate system ($x$,$y$), there is no rotation of the vehicle if the resulting net torque vanishes; it may easily be shown that the 
condition for no rotation is:
\begin{equation} \label{eq:CondNoRot}
\sin \alpha (F_1+F_2-F_3-F_4) (L_x \sin \alpha - L_y \cos \alpha) = 0 \, \, \, .
\end{equation}
This equation admits three solutions, one of which (namely, $\sin \alpha = 0$) is trivial and uninteresting from the point of view of a practical application. However, the two remaining conditions are worth noting. Evidently, in case of right/left 
balance of the forces (i.e., when $F_1+F_2=F_3+F_4$), the net torque (in the standard configuration of the Mecanum wheels) vanishes. The last condition is the most interesting one: if the track and the wheelbase are chosen in such a way as to satisfy 
the condition $\tan \alpha=L_y/L_x$, the resulting net torque always vanishes \emph{irrespective of the magnitudes of the applied forces and of the direction of the rotation of the Mecanum wheels}! (Given that both $L_x$ and $L_y$ are positive, the 
last equation becomes relevant only when $\alpha \in (0,\pi/2)$.) This result must be borne in mind in case that the rotational effects must be suppressed. Finally, one may show that, if the condition $F_1+F_2=F_3+F_4$ is fulfilled, the vehicle will 
move along a straight path, the direction of which will involve the angle $\tan^{-1} (\frac{(F_3+F_4) \tan \alpha}{F_2-F_3})+\phi(0)$ with respect to the $X$ axis of the laboratory coordinate system.

\subsection{\label{sec:Formalism4}Derivation of the solution}

Let us assume that, at time $t_1$, the known quantities are:
\begin{itemize}
\item the position vector of the CM of the vehicle $\overrightarrow{R_{CM}}$,
\item the velocity of the CM of the vehicle $\overrightarrow{V_{CM}}$,
\item the orientation angle $\phi$,
\item the angular velocity $\omega$, and
\item the tractive forces $\vec{F}_i$ for each wheel.
\end{itemize}

The output at time $t_1 + \Delta t$ comprises:
\begin{itemize}
\item the position vector of the CM of the vehicle,
\item the velocity of the CM of the vehicle,
\item the orientation angle, and
\item the angular velocity.
\end{itemize}

Compared to systems using standard wheels, it is more complex to include the resistive effects in the general motion of a Mecanum-wheeled vehicle. In the former case, the wheels move in a concerted manner, thus creating either forward or backward 
overall motion; a vehicle using Mecanum wheels is bound to behave differently. Not only can the wheels be set to rotate at different angular velocities, but they might also be set to rotate against each other. As a result, the general state of 
motion of each wheel is best described as `rolling with sliding'; this implies that, in the general case, the sliding-friction coefficient appears to be the appropriate one to use in the expressions modelling the motion. On the other hand, if the 
motion is concerted (i.e., when the wheels are set to move in the same direction, at equal velocities), the coefficient of rolling resistive friction should better be used. To enable the inclusion in the description of the motion of the effects 
pertaining to friction and simultaneously retain simplicity, the following scheme is proposed.
\begin{enumerate}
\item Loop over the four wheels:
\begin{itemize}
\item Estimate $\vec{V}_i$ via Eq.~(\ref{eq:VI}) and its projection along the $\hat{u}_{i,p}$ direction, i.e., the $V_{i,p}$ component entering the evaluation of the resistive friction exerted on the wheel by the floor.
\item Estimate the component of the thrust along the $\hat{u}_{i,p}$ direction (i.e., $F_{i,p}$).
\item If $\mid V_{i,p} \mid$ exceeds a tolerance limit (denoted in the flowchart below by the quantity $\epsilon$), then the resistive force is given by the sum of the rolling- and viscous-friction contributions 
($\vec{f}_{i,p} = - sgn(V_{i,p}) \mu_r N_i \hat{u}_{i,p} - \mu_v N_i \vec{V}_{i,p}$) if $V_{i,p}$ and $F_{i,p}$ are of the same sign, whereas by the sum of the sliding- and viscous-friction contributions 
($\vec{f}_{i,p} = - sgn(V_{i,p}) \mu_s N_i \hat{u}_{i,p} - \mu_v N_i \vec{V}_{i,p}$) if $V_{i,p}$ and $F_{i,p}$ are of opposite sign; if, on the other hand, $\mid V_{i,p} \mid < \epsilon$ but 
$\mid F_{i,p} \mid > \mu_r N_i$, then the friction is overcome by the thrust and the condition for the motion of the wheel along the $\hat{u}_{i,p}$ direction is fulfilled (the motion will be allowed 
within the current period of time $\Delta t$). Of course, if the thrust has not overcome the friction, the wheel will not move within the time interval $\Delta t$.
\item Redefine $F_{i,p}$ by taking the resistive forces $f_{i,p}$ into account. The following flowchart describes this simple procedure, applied to each wheel $i$ of the vehicle.
\end{itemize}
\item Evaluate the components $F_x$ and $F_y$ from the forces $\vec{F}_{i,p}$ (by considering the projections of the unit vectors $\hat{u}_{i,p}$ onto the $x$ and $y$ directions); evaluate $F_X$ and $F_Y$ from the matrix equation (\ref{eq:FLAB}).
\item Use expressions (\ref{eq:VCM}-\ref{eq:Phi}) to obtain the output values of the relevant physical quantities at time $t_1 + \Delta t$.
\end{enumerate}

\begin{tikzpicture}[scale=2, node distance = 5cm, auto]
\node [decision] (d1) {$\mid V_{i,p} \mid > \epsilon$?};
\node [decision, below of=d1] (d2) {$V_{i,p} \cdot F_{i,p}>0$?};
\node [decision, right of=d1] (d3) {$\mid F_{i,p} \mid > \mu_r N_i$?};
\node [block, below of=d2] (b1) {$\vec{F}_{i,p} \leftarrow \vec{F}_{i,p} - sgn(V_{i,p}) \mu_r N_i \hat{u}_{i,p} - \mu_v N_i \vec{V}_{i,p}$};
\node [block, right of=d2] (b2) {$\vec{F}_{i,p} \leftarrow \vec{F}_{i,p} - sgn(V_{i,p}) \mu_s N_i \hat{u}_{i,p} - \mu_v N_i \vec{V}_{i,p}$};
\node [block, above of=d3] (b3) {$\vec{F}_{i,p} \leftarrow \vec{F}_{i,p} - sgn(F_{i,p}) \mu_r N_i \hat{u}_{i,p}$};
\node [block, right of=d3] (b4) {$\vec{F}_{i,p} \leftarrow \vec{0}$};
\path [line] (d1) -- node {yes} (d2);
\path [line] (d2) -- node {yes} (b1);
\path [line] (d2) -- node {no} (b2);
\path [line] (d1) -- node {no} (d3);
\path [line] (d3) -- node {yes} (b3);
\path [line] (d3) -- node {no} (b4);
\end{tikzpicture}

It is convenient to place the CM of the vehicle at $t=0$ at the origin of the laboratory coordinate system, at rest; therefore, $\overrightarrow{R_{CM}} (0)=\vec{0}$ and $\overrightarrow{V_{CM}} (0)=\vec{0}$. Additionally, $\vec{\phi} (0)=\vec{0}$ 
and $\vec{\omega} (0)=\vec{0}$. One may then start to apply the tractive forces to the wheels, perhaps using a linear model $F_i \frac{t}{T}$ for $t \leq T$, and the nominal values $F_i$ for $t>T$; a reasonable value for $T$ may be a few tenths of 
$1$s. Concerning the time increment $\Delta t$, it must be sufficiently small to ensure the accuracy of the output. Assuming that the execution time is not an issue, one might put forth a scheme of subdividing the original time increment $\Delta t$ 
into segments $\delta t$, obtain the relevant physical quantities (at $t_1+\Delta t$) for several $\delta t$ values, fit the values with a (simple) polynomial, and (finally) obtain the results at $t_1+\Delta t$ using the extrapolated values to $\delta t=0$.

\subsection{\label{sec:Formalism5}Simple examples of motion}

The implementation of the formalism, developed in Sections \ref{sec:Formalism1}-\ref{sec:Formalism4}, led to the results of Table \ref{tab:OM} for a few simple motions; a similar table may be found in Ref.~\cite{tv}~\footnote{One must bear in mind that 
Ref.~\cite{tv} differs from this work in some definitions, e.g., in the sign convention of the $\omega_i$'s.}.

A change of orientation of the vehicle $y$ axis (with respect to the laboratory coordinate system) may be caused not only via the application of different thrusts (or rotations) on the Mecanum wheels, but also by shifting the CM of the vehicle (away from 
its geometrical centre). This is evident after inspecting Eq.~(\ref{eq:TQ}); if $\overrightarrow{r_{CM}} \neq \vec{0}$, a (non-zero) torque is generated as soon as the vectors $\overrightarrow{r_{CM}}$ and $\vec{F}$ are not aligned (or anti-aligned). The 
results of Table \ref{tab:OM} have been obtained with $\overrightarrow{r_{CM}}=\vec{0}$.

\begin{figure}
\begin{center}
\includegraphics [width=14cm] {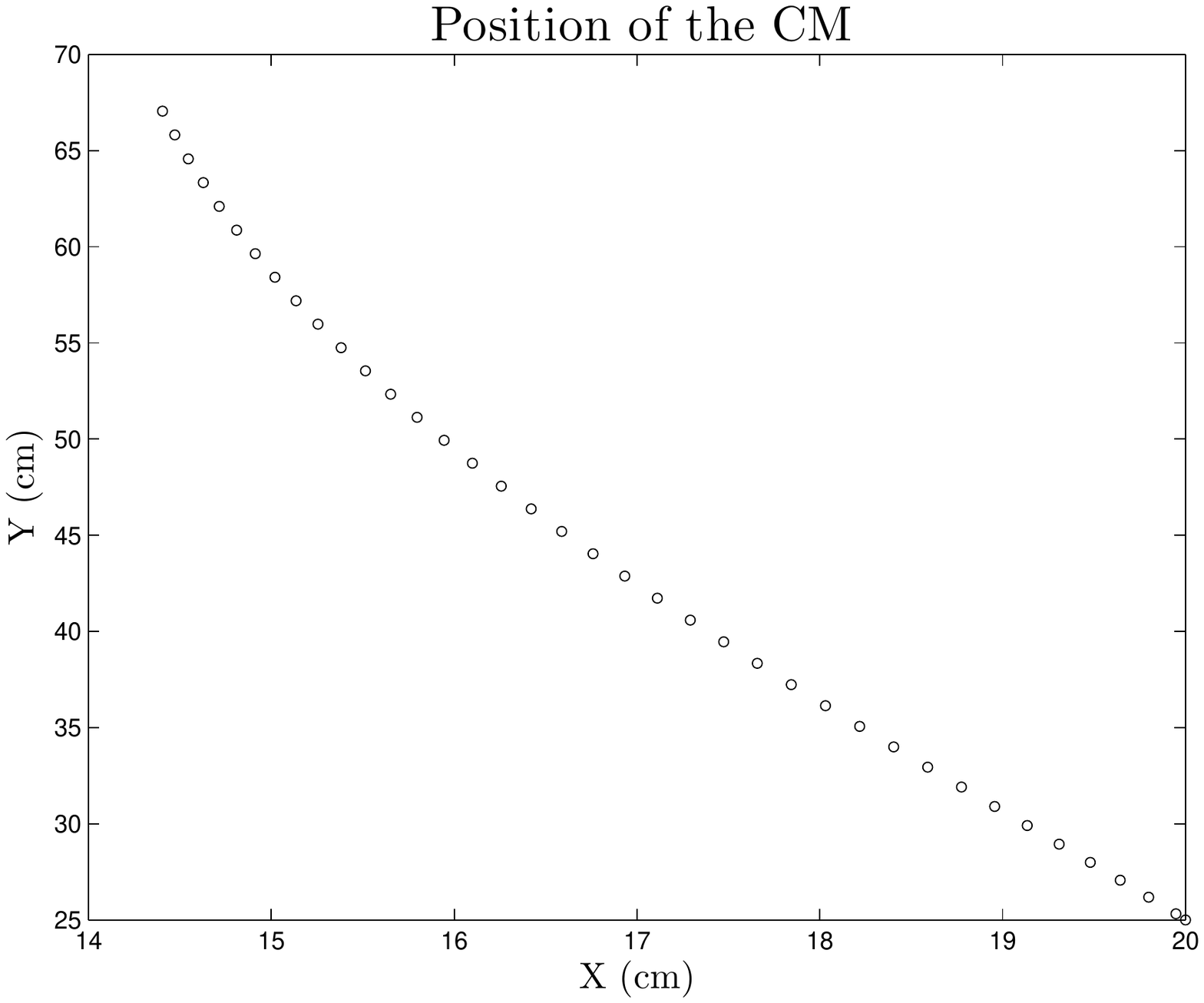}
%\vspace{-6cm}
\caption{\label{fig:Bogie7}The position coordinates of the centre of mass (CM) of a Mecanum-wheeled vehicle in the laboratory coordinate system. The parameters generating this motion are given in Section \ref{sec:Formalism5}.}
\end{center}
\end{figure}

\begin{figure}
\begin{center}
\includegraphics [width=14cm] {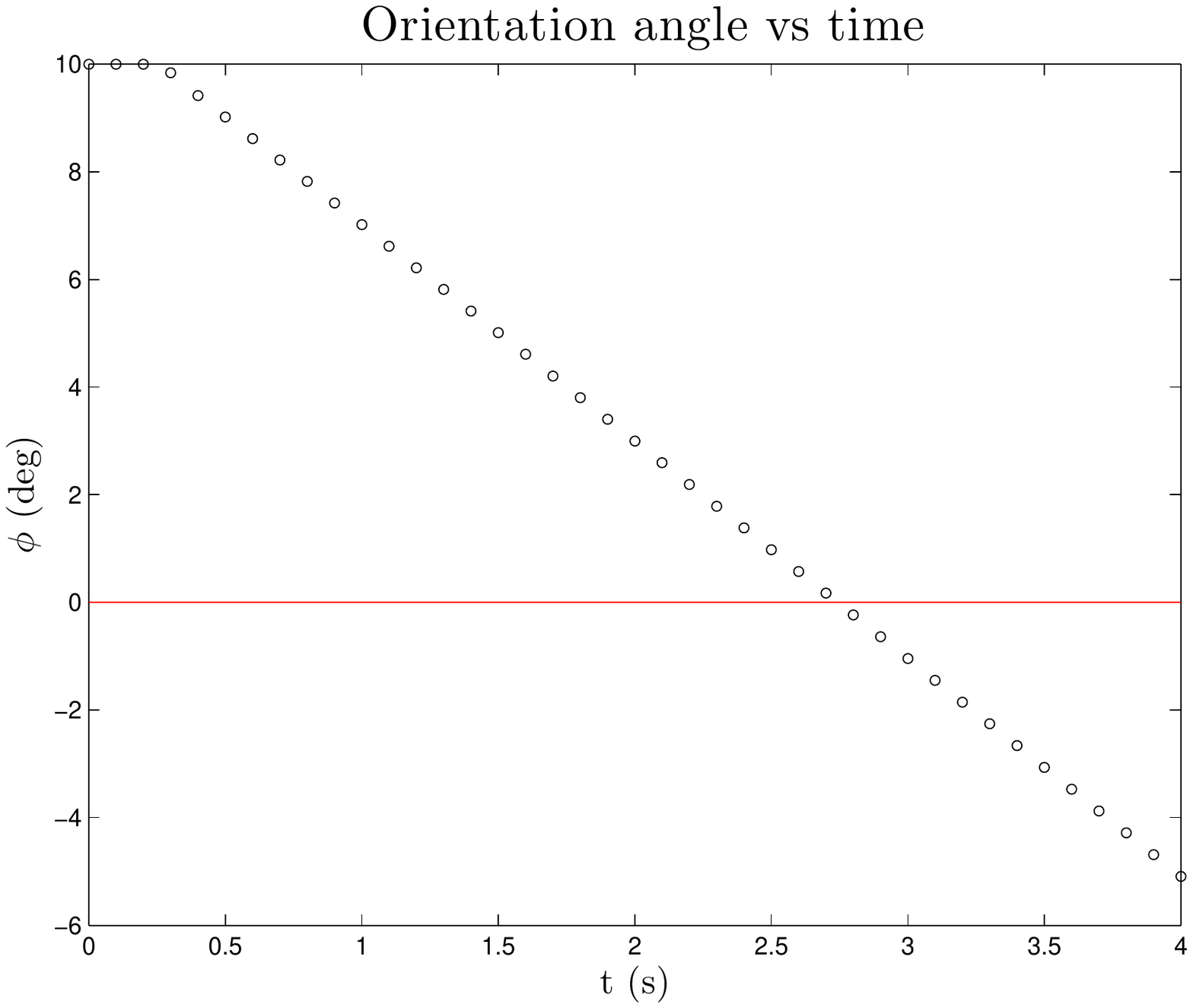}
%\vspace{-6cm}
\caption{\label{fig:Bogie8}The time dependence of the orientation angle $\phi$. The parameters generating this motion are given in Section \ref{sec:Formalism5}.}
\end{center}
\end{figure}

\begin{figure}
\begin{center}
\includegraphics [width=14cm] {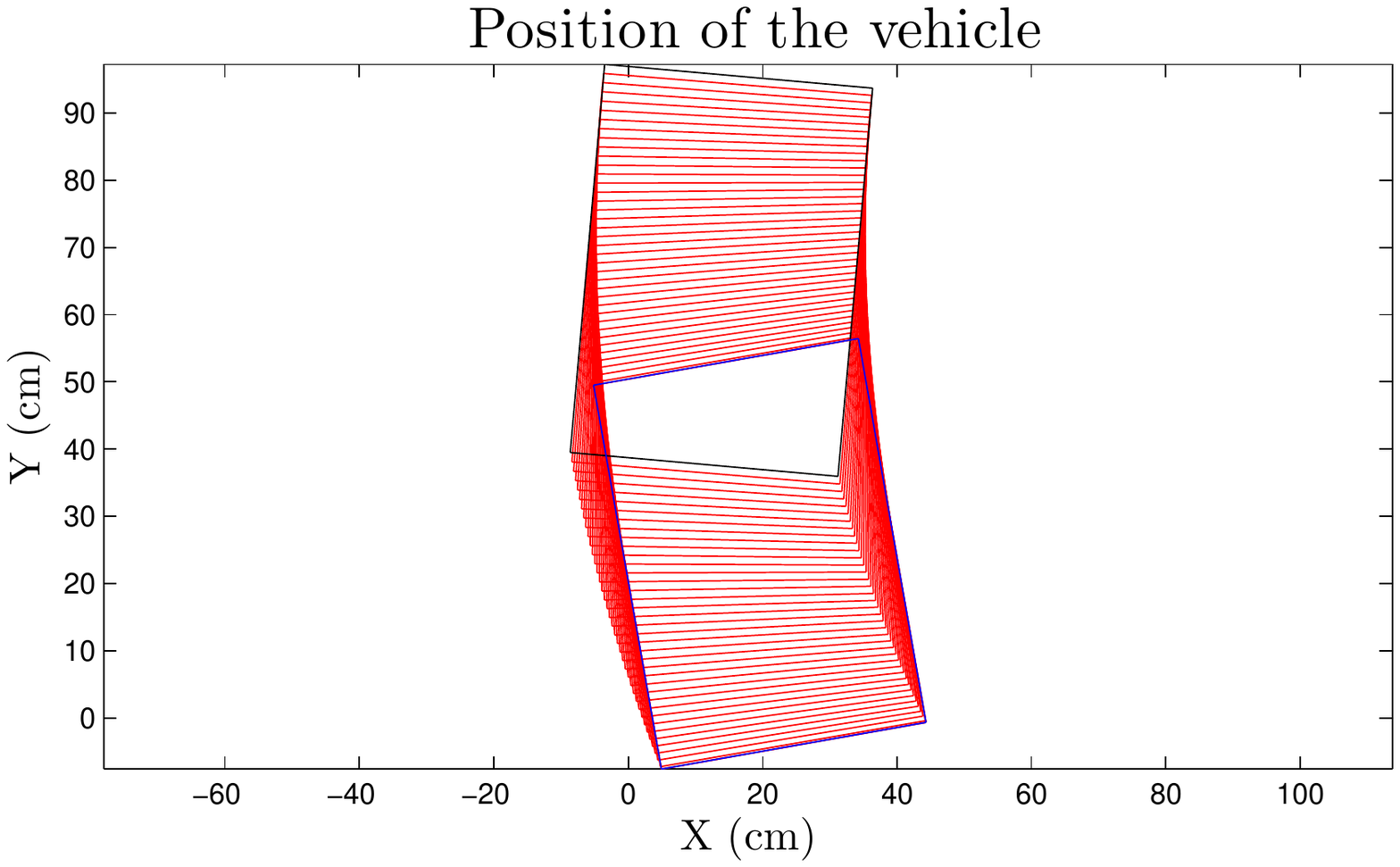}
%\vspace{-6cm}
\caption{\label{fig:Bogie9}The position and orientation of a Mecanum-wheeled vehicle in the laboratory coordinate system. The final position (i.e., at $t=4$s) corresponds to the frame shown in black, the starting one in blue. The parameters generating this 
motion are given in Section \ref{sec:Formalism5}.}
\end{center}
\end{figure}

One case of motion of a $16.6$kg vehicle is displayed in Figs.~\ref{fig:Bogie7}-\ref{fig:Bogie9}, over a period of $4$s, in $100$ms steps. The parameters generating this motion are (see next section): $\nu_1=10$, $\nu_2=12$, $\nu_3=14$, $\nu_4=16$ revolutions 
per minute; $sgn(\omega_1)=sgn(\omega_2)=sgn(\omega_3)=sgn(\omega_4)=+1$ (`++++' in the notation of Table \ref{tab:OM}); $N=250$kp; $L_x=40$, $L_y=58$cm; $\overrightarrow{r_{CM}}=(5 \hat{u}_x + 5 \hat{u}_y)$mm; $\alpha=130^\circ$; $\mu_s=0.51$, $\mu_r=0.084$; 
$\mu_v=0.15$s/m. The starting position in the laboratory coordinate system ($X$, $Y$, $\phi$) is ($20$cm, $25$cm, $10^\circ$). The thrusts are applied to the wheels linearly over $T=300$ms; the nominal thrust values are assumed to have been reached at $t=T$ 
(see last paragraph of the previous section).

In the present work, the assumption is that the load is applied at \emph{one} specific point of the vehicle surface. As a result, the solution (extracted for the rotation of the vehicle) is, generally speaking, not very realistic; in reality, the application 
of the load involves the contact of two surfaces. Consequently, the amount the vehicle rotates in the ideal case is expected to exceed the actual amounts obtained in realistic cases. This is due to two reasons. Firstly, if the rigid object (which is hard to 
move in the attached coordinate system ($x$,$y$)) is placed onto the Mecanum-wheeled vehicle, part of the driving torque will be `wasted' in balancing the frictional torque, in analogy to the trade-off between tractive and resistive forces in translational 
motion. Secondly, the value of the moment of inertia $I$ in Eqs.~(\ref{eq:N2}, \ref{eq:Omega}, \ref{eq:Phi}) will have to include an unknown (positive) contribution pertaining to the geometrical characteristics of the object comprising the load. One way to 
solve part of the former problem is to allow the rotation of the vehicle only if the driving torque exceeds a specific value, and subtract a (perhaps, different) contribution from the driving torque after the rotation of the vehicle has been initiated (i.e., 
distinguish between static and dynamic frictional torques); without specific information on the characteristics of the object responsible for the load, it is not possible to obtain a solution to the latter problem.

\section{\label{sec:ER}Validation of the model}

\subsection{\label{sec:Experiments}Experiments}

A number of tests have been carried out in order to validate the model of Section \ref{sec:Method} and to examine whether the values of its parameters could be determined from experimental data; the last of these tests (performed on the concrete floor 
of the IMS laboratory on October 3, 2012) was documented. The results of the previous tests had already shown that:
\begin{itemize}
\item the vehicle was controllable, in fact, equally-well controllable over straight paths set at angles between $0$ and $90^\circ$ with respect to the $X$ axis of the laboratory coordinate system (initial condition: $\phi(0)=0^\circ$), and
\item the angular uncertainty in its motion did not exceed $1^\circ$ (root-mean-square value of the difference between expected and actual directions).
\end{itemize}
For the documented data taking, a number of angular-velocity combinations (as well as rotational directions), resulting in a motion of the $16.6$kg prototype within the first quarter-circle (i.e., $X(t)>0$ and $Y(t)>0$), were applied to the wheels of the 
vehicle; the coordinates of the end points (of the trajectories) were measured relative to the walls of the test room, using a laser distance meter (Leika DistoTM A2 Laser). The motion was interrupted when the distance between the starting and final 
positions of the right-rear corner of the vehicle was about $2$m (due to imperfection of the floor, the motion was interrupted earlier, at distances between $1.6$ and $1.7$m, below $20^\circ$). The details on these data may be found in Table \ref{tab:Test}. 
(The inclination angle $\alpha$ in the prototype is $135^\circ$.)

The friction coefficients were also measured in the two following cases: a) the brakes were applied unto the wheels and b) the wheels were allowed to rotate freely. The former measurement yields the coefficient of the sliding friction, whereas the latter 
one fixes the coefficient of the rolling friction. The results were: $\mu_s = 0.51 \pm 0.03$ and $\mu _r = 0.084 \pm 0.006$.

The experimental data (a total of $M=32$ measurements of the coordinates of the end points of straight paths, as well as an equal amount of data pertaining to the time needed in order to accomplish each motion) were submitted for optimisation. Two additional 
parameters were introduced (denoted as $A$ and $B$), to associate the angular velocity $\omega_i$ of each wheel with the tractive force $F_i$ acting on that wheel~\footnote{In order to determine the motion of the vehicle, it is necessary to associate the 
angular velocity $\omega_i$ of each wheel with the tractive force $F_i$ acting on that wheel. In the future, this relation must be established after analysing dedicated experimental data, corresponding to the motors, the materials involved, and the geometrical 
and physical characteristics of the specific wheels used in the vehicle.}; a simple linear model was put forth: $F_i \, = \, A \, R \, \omega_i \, + \, B$, where $R$ stands for the radius of the wheels used in the prototype ($R=5$cm).

For a given set of parameter values, the motion of the vehicle was determined until the measured end point was `boxed'. An estimation of the shortest distance $d_i$ of the resulting trajectory to the measured data point was next obtained. The uncertainty in 
this determination $\delta d_i$, to be assigned to each experimental value, was assumed proportional to the distance between the starting and final positions (of the reference point of the vehicle); the assumed angular uncertainty of $1^\circ$ is translated 
into a spatial uncertainty of about $3.5$cm over the distance of $2$m. As the distances between the starting and final positions were kept close to $2$m, the assigned statistical uncertainties of the input data (above $20^\circ$) were almost equal for the entire 
set of measurements. The optimisation was achieved via the minimisation of the $\chi^2$ function, defined by the formula:
\begin{equation} \label{eq:MF}
\chi^2=\sum_{i=1}^{M} \Big\{ \big( \frac{d_i}{\delta d_i} \big)^2 + \big( \frac{t_i^{exp}-t_i^{th}}{\delta t_i} \big) ^2 \Big\} \, \, \, ,
\end{equation}
where $t_i^{exp}$ and $t_i^{th}$ denote the measured and fitted (estimated with the model) times corresponding to the particular motion; the uncertainties $\delta t_i$ were `globally' set to $2$s (which may have been an overestimation of the true uncertainties). The 
free parameters were varied until the description of the input experimental data was optimised (i.e., until the $\chi^2$ minimum was obtained). Given that two physical quantities (i.e., the distance $d_i$ and the time $t_i^{exp}$) per case were optimised, 
the total number of input data points is $2 M=64$.

For the purpose of the optimisation, the standard MINUIT package \cite{james} of the CERN library was used, namely the C++ version of the library \cite{jw}. Extensive information on this software package may be obtained online \cite{m2}. The latest release 
of the code is available from Ref.~\cite{m2onln}. Version $5.28.00$ of Minuit2 was incorporated in the analysis framework. Each optimisation was achieved on the basis of the (robust) SIMPLEX-MINIMIZE-MIGRAD-MINOS chain.

The first fits to the experimental data, treating all five model parameters (i.e., the three coefficients of friction, $A$, and $B$) as free, revealed the presence of strong correlations. To mitigate these correlations, a decision had to be made as to 
which parameters could be fixed from external sources. Given that the coefficients of the sliding and rolling friction had already been obtained from direct measurements, it was decided to use these two results and vary only $\mu_v$, $A$, and $B$ in the 
fits. Due to the largeness of the correlations, a cautious stepwise approach during the next steps of the optimisation phase was followed. Apart from occasional fixing and releasing some of the model parameters (thus enabling the examination of the 
sensitivity of the results to the variation of the model parameters, as well as the assessment of the performance of the optimisation algorithms in the problem), the optimisation scheme was also run including and excluding the measurements relating to the 
duration of the motion; the results of this test demonstrated that the problems encountered in the data analysis did not originate from the inclusion of the time measurements in the fit. However, despite all these efforts, no satisfactory results were 
obtained; a number of fits failed to yield meaningful output (e.g., the Hessian matrix could not be evaluated) as the optimisation algorithms were trapped in spurious $\chi^2$ minima. (A dependence of the results of the fit on the initial guess of the 
model-parameter values was also observed.) On almost all occasions, the free (unconstrained) fits drifted as a result of the existing correlations. In order to cope with this behaviour, it was subsequently decided to use available information on some technical details 
of the hardware being used in the tests, and impose appropriate constraints on the model parameters (e.g., pertaining to the maximal power and torque of the motors used in the prototype), on top of the obvious physically-imposed constraints (e.g., positivity 
of the friction coefficients, etc.). Despite these actions, the tendency of the fits to drift persisted.

It currently seems that the only way to obtain a stable solution was to perform the fits at fixed $\mu_v$ values, taken from a representative interval. The question of what `representative' means is better answered after considering the viscous friction as the 
velocity-dependent component of the (varying with the velocity) sliding/rolling friction force~\footnote{In other words, whichever the dependence of the sliding/rolling friction on the relative velocity of the two surfaces in contact, it will be modelled via a 
linear function; given the smallness of the typical velocities in the problem, this approximation is expected to be good.}. From this viewpoint, one could set limits for the allowed variation of the sliding-friction coefficient between two reference velocities, 
say, between $0$ and $1$m/s (arbitrary choice). It should suffice to assume a variation of $100 \%$ (of the sliding-friction coefficient, assumed to be velocity-dependent) between these two reference velocities. In that case, the $100 \%$ variation of the 
coefficient is equivalent to varying the parameter $\mu_v$ from $0$ to the value of $\mu_s$ (in s/m units). Since each optimisation (at a fixed $\mu_v$ value) involves two free parameters, the number of degrees of freedom in the fits was ${\rm NDF}=64-2=62$.

\subsection{\label{sec:Results}Results}

The results of the optimisation for fixed values of the parameter $\mu_v$ from ($0.05$ to $0.50$s/m, with a step of $0.05$s/m) are shown in Table \ref{tab:OptRes}. Inspection of the table (and after trivial numerical operations on the data it contains) 
provides an answer to the question of the failure of the optimisation scheme in yielding meaningful results. The $\chi^2$ function appears to be monotonic with $\mu_v$ (at least for the $\mu_v$ interval of interest, as chosen herein). This behaviour of the 
minimisation function prevents the optimisation algorithms from terminating successfully. The fit drifts in search of the $\chi^2$ minimum, which (judging from the corresponding plot) is nowhere in the vicinity of the interval of representative $\mu_v$ values. 
In any case, Table \ref{tab:OptRes} will surely be useful when simulating the motion of the vehicle in the general case; the user may choose any of the shown ($\mu_v$,$A$,$B$) combinations, exempting perhaps only the first row (i.e., the record 
corresponding to $\mu_v=0.05$s/m).

In view of these results, it makes sense to restrict the analysis to a reasonable interval of $\mu_v$ values, e.g., $0.10 \leq \mu_v \leq 0.20$s/m. In that case, the average of the three minimal $\chi^2$ values is about $105.9$ for $62$ degrees of freedom 
in the fit, i.e., about $1.7$ units (on average) per degree of freedom; this reduced-$\chi^2$ value is reasonable. A more careful analysis of the results of the fits indicated that the motion was more controllable for small (say, below about $35^\circ$) 
and large (above about $75^\circ$) angles; the data in these regions were better reproduced (than those outside) with the model of Section \ref{sec:Method}. The worst cases in the reproduction of the coordinates of the end points occurred around $60^\circ$, 
where the reduced-$\chi^2$ values were as large as $6.7$, showing a deterioration from large to small angular velocity. This might be due to the influence of effects which have not been taken into account in the model of Section \ref{sec:Method} (e.g., see 
Ref.~\cite{vt}).

\begin{figure}
\begin{center}
\includegraphics [width=14cm] {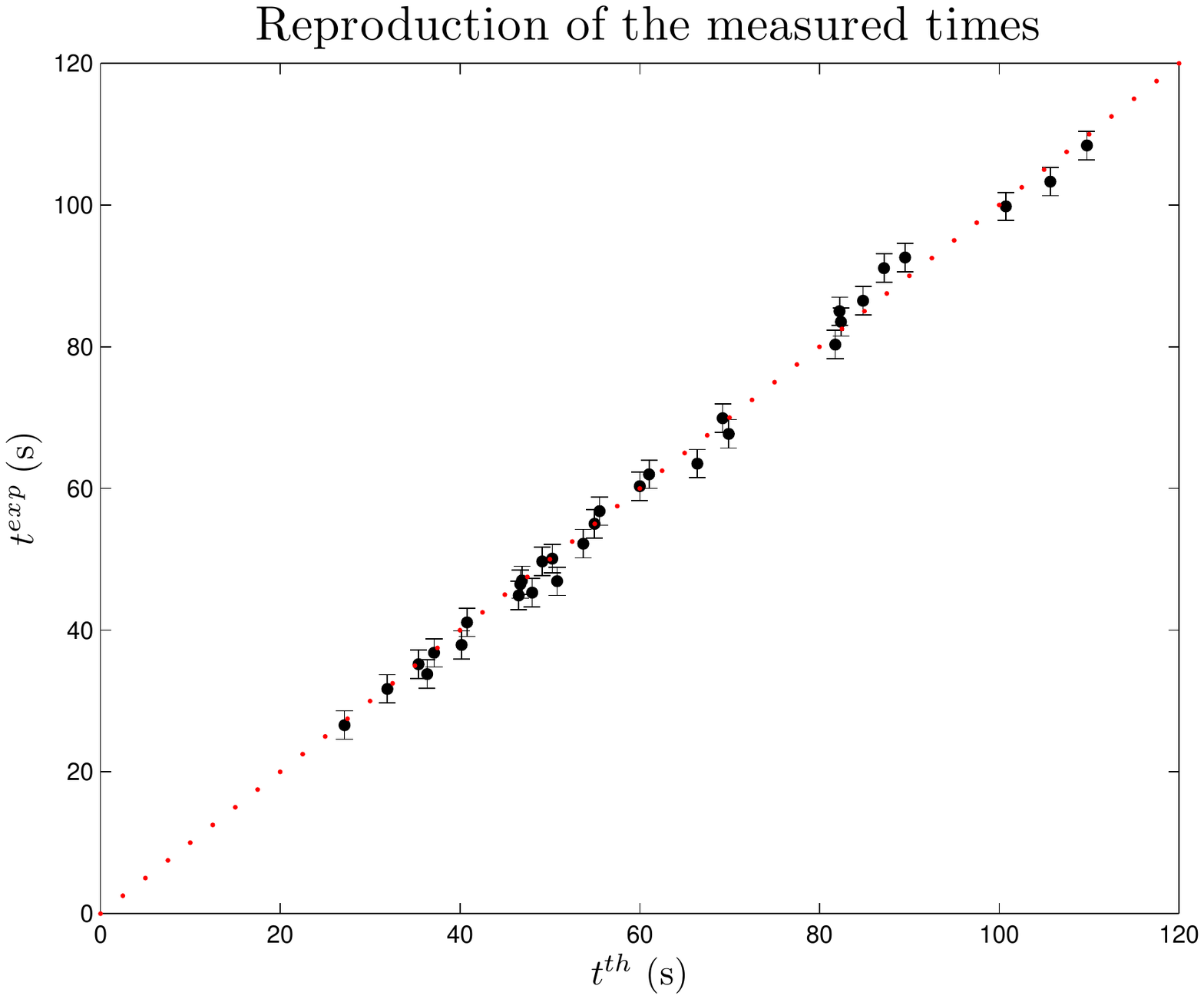}
%\vspace{-6cm}
\caption{\label{fig:TimeMeasurementReproduction}Reproduction of the measured times needed for the accomplishment of motion for the $32$ test cases of Table \ref{tab:Test}. The dotted line indicates equal measured ($t^{exp}$) and fitted ($t^{th}$) times.}
\end{center}
\end{figure}

The reproduction of the time duration of the motion (see Fig.~\ref{fig:TimeMeasurementReproduction}) was found superior to that of the coordinates of the end points of the trajectories. The $\chi^2$ value, corresponding to the reproduction of the `temporal' 
data, was equal to only $28.0$ units for $32$ measurements. The linear-correlation coefficient between measured and fitted values exceeded $0.9975$.

\subsection{\label{sec:Comments}Further comments}

It is true that another approach could have been followed in the optimisation phase. For instance, the trajectory could be determined for a time equal to the one measured in the experiment; the coordinates of the end points of the resulting trajectory could 
then be compared directly to those obtained experimentally. In this scheme, only the first term of the minimisation function of Eq.~(\ref{eq:MF}) is relevant. One could also introduce the time uncertainty $\delta t_i$, by appropriately translating it into a 
spatial uncertainty and combining it (e.g., quadratically) with the (assigned) spatial uncertainties $\delta d_i$ of the data points. In principle, this approach and the one followed herein should lead to similar results.

The measurements obtained during the experiment rely on the determination of the distances of the reference point of the prototype to the walls of the test room. Concerning these measurements, the assumption is that the planes of these walls are perpendicular 
to each other (Cartesian laboratory coordinate system). Unfortunately, a deviation of about $1^\circ$ was observed after simply placing a right-angle ruler at the corner of the two walls. Therefore, it makes sense to take the numbers of Table \ref{tab:Test} 
with a grain of salt. Equivalently, one could accept these values (and the bias they might introduce into the analysis), but augment the input uncertainties of the data points; the latter strategy was adopted herein. (The use of a laser scanner, as in 
Ref.~\cite{vt}, would be a considerable improvement at this point.)

The model of Section \ref{sec:Method} was put forth in order to understand the motion of one vehicle using Mecanum wheels, so that a more complex system (e.g., a swarm of such vehicles) be efficiently studied. In this context, it is not relevant whether all 
physical effects have analytically been included in the proposed model, as long as the influence of any missing pieces is captured by the model parameters, which (in this case) lose their clear physical interpretation and become effective. Finally, by no means 
should the tests performed herein be considered complete; to thoroughly test the model of Section \ref{sec:Method}, one would need to develop a test platform similar to the one shown in Fig.~$7$ of Ref.~\cite{vt}.

\section{\label{sec:Conclusions}Conclusions}

In the present paper, given are the details of the modelling of the omni-directional, two-dimensional motion of a vehicle using Mecanum wheels. The model is based on simple Mechanics and includes the effects of rolling, sliding, and viscous friction. The motion 
is obtained in the general case, i.e., for arbitrary thrusts applied to the wheels of the vehicle.

The three friction coefficients comprise the parameters of the model. Two additional parameters were introduced at the present stage, in order to account for the dependence of the tractive force (applied to each wheel) on the corresponding angular velocity; in 
the future, these parameters could either be fixed from dedicated experimental data or be directly derived from the detailed technical specifications of the motors, as well as the geometrical and physical properties of the wheels.

The model was confronted with recently-acquired experimental data, comprising the coordinates of the end positions of one reference point of the vehicle for straight paths in the first quarter-circle, as well as the times needed in order to achieve each 
motion. Given the strong correlations, which were present among the model parameters in the optimisation of the description of the experimental data, two of the friction coefficients (i.e., of rolling and sliding friction) were fixed from direct measurements. 
This action, however, did not suppress the correlations. To obtain meaningful results from the experimental data, the fits had to be performed at fixed values of the viscous-friction coefficient, taken from an interval which appears to be reasonably broad.

The present work may serve as the basis for the simulation of the motion of one vehicle (for arbitrary thrusts, inclination angle of the axles of the rollers, loads, and offsets of the centre of mass of the vehicle), as well as of the concerted motion of a 
swarm of such vehicles. Obviously, further work is needed both on the theoretical, as well as on the experimental aspects in this topic.

\begin{ack}
The author acknowledges a useful exchange of electronic mail with M. de Villiers. The author also acknowledges helpful discussions with the members of the AVERT team at ZHAW: A.N. Chand, M.D. Coray, G.C. Deshmukh, S. Fluck, N.A. Haggenmacher, C.M. Henschel, 
R.A. H{\"u}ppi, J. Kr{\"u}si, V. Meiser, D. {\v S}eatovi{\' c}, and J. Wirth. The data of the present paper were obtained in collaboration with C.M. Henschel, who also developed the software for steering the prototype used in the performed experiments. This 
study is part of the Autonomous Vehicle Emergency Recovery Tool (AVERT) project (project number: 285092), funded by the EU.
\end{ack}

\newpage
\begin{table}[h!]
{\bf \caption{\label{tab:OM}}}A few cases of simple motion of a Mecanum-wheeled vehicle; a similar table may be found in Ref.~\cite{tv}. CW denotes `clockwise', CCW `counter-clockwise'. The signs correspond to the quantities $sgn(\omega_i)$; $0$ 
indicates no rotation. The cardinal directions correspond to a top view of the motion of the vehicle, i.e., E is the direction of positive $X$ (at constant $Y$). The same thrust value was applied to the wheels. The orientation of the rollers axles 
was set to $45^\circ$.
\vspace{0.6cm}
\begin{center}
\begin{tabular}{|c|c|c|c|c|}
\hline
Wheel $1$ & Wheel $2$ & Wheel $3$ & Wheel $4$ & Direction of motion \\
\hline
$+$ & $+$ & $+$ & $+$ & N \\
$0$ & $+$ & $+$ & $0$ & N \\
$+$ & $0$ & $0$ & $+$ & N \\
\hline
$-$ & $-$ & $-$ & $-$ & S \\
$0$ & $-$ & $-$ & $0$ & S \\
$-$ & $0$ & $0$ & $-$ & S \\
\hline
$+$ & $-$ & $+$ & $-$ & W \\
$+$ & $-$ & $0$ & $0$ & W \\
$0$ & $0$ & $+$ & $-$ & W \\
\hline
$-$ & $+$ & $-$ & $+$ & E \\
$-$ & $+$ & $0$ & $0$ & E \\
$0$ & $0$ & $-$ & $+$ & E \\
\hline
$+$ & $0$ & $+$ & $0$ & NW \\
$-$ & $0$ & $-$ & $0$ & SE \\
$0$ & $+$ & $0$ & $+$ & NE \\
$0$ & $-$ & $0$ & $-$ & SW \\
\hline
$+$ & $+$ & $-$ & $-$ & CW Rotation \\
$-$ & $-$ & $+$ & $+$ & CCW Rotation \\
\hline
\end{tabular}
\end{center}
\end{table}

\newpage
\begin{table}[h!]
{\bf \caption{\label{tab:Test}}}Details on the input data acquired on the concrete floor of the IMS laboratory on October 3, 2012. These values are analysed in Section \ref{sec:ER}. The quantity $\nu$ denotes the revolution rate (number of revolutions 
of the relevant wheel per unit time); in all cases, $\nu_3=\nu_1$ and $\nu_4=\nu_2$.
\vspace{0.6cm}
\begin{center}
\begin{tabular}{|c|c|c|c|c|}
\hline
$t_i^{exp}$ (s) & $X_i^{exp}$ (m) & $Y_i^{exp}$ (m) & $\nu_1$ (min$^{-1}$) & $\nu_2$ (min$^{-1}$) \\
\hline
$26.6$ & $1.5770$ & $0.2615$ & $14.55$ & $-10.19$ \\
$35.2$ & $1.6180$ & $0.2675$ & $11.25$ & $-7.88$ \\
$49.7$ & $1.6190$ & $0.2755$ & $7.95$ & $-5.57$ \\
$85.0$ & $1.6320$ & $0.2665$ & $4.65$ & $-3.26$ \\
\hline
$31.7$ & $1.6130$ & $0.5855$ & $14.55$ & $-6.78$ \\
$41.1$ & $1.6440$ & $0.5805$ & $11.40$ & $-5.32$ \\
$56.8$ & $1.6500$ & $0.5755$ & $8.25$ & $-3.85$ \\
$91.1$ & $1.6380$ & $0.5785$ & $5.10$ & $-2.38$ \\
\hline
$36.8$ & $1.6270$ & $0.9525$ & $14.55$ & $-3.90$ \\
$47.0$ & $1.6500$ & $0.9715$ & $11.55$ & $-3.09$ \\
$62.0$ & $1.6440$ & $0.9585$ & $8.70$ & $-2.33$ \\
$92.6$ & $1.6550$ & $0.9575$ & $5.85$ & $-1.57$ \\
\hline
$50.1$ & $1.8820$ & $1.6515$ & $14.55$ & $-1.27$ \\
$55.0$ & $1.7440$ & $1.5405$ & $12.30$ & $-1.08$ \\
$69.9$ & $1.8430$ & $1.6165$ & $10.20$ & $-0.89$ \\
$86.5$ & $1.7970$ & $1.6025$ & $8.10$ & $-0.71$ \\
\hline
$46.5$ & $1.4120$ & $1.8425$ & $14.55$ & $1.27$ \\
$60.3$ & $1.4160$ & $1.8515$ & $11.25$ & $0.98$ \\
$83.5$ & $1.4400$ & $1.8915$ & $8.25$ & $0.72$ \\
$123.4$ & $1.5400$ & $2.0485$ & $6.00$ & $0.53$ \\
\hline
$44.9$ & $1.0980$ & $2.1195$ & $14.55$ & $3.90$ \\
$52.2$ & $1.0750$ & $2.0855$ & $12.30$ & $3.30$ \\
$80.3$ & $1.3420$ & $2.6365$ & $10.05$ & $2.69$ \\
$99.8$ & $1.2810$ & $2.5515$ & $7.80$ & $2.09$ \\
\hline
\end{tabular}
\end{center}
\end{table}

\newpage
\begin{table*}
{\bf Table \ref{tab:Test} continued}
\vspace{0.6cm}
\begin{center}
\begin{tabular}{|c|c|c|c|c|c|c|}
\hline
$t_i^{exp}$ (s) & $X_i^{exp}$ (m) & $Y_i^{exp}$ (m) & $\nu_1$ (min$^{-1}$) & $\nu_2$ (min$^{-1}$) \\
\hline
$37.9$ & $0.6710$ & $2.0785$ & $14.55$ & $6.79$ \\
$46.9$ & $0.6640$ & $2.1185$ & $11.55$ & $5.39$ \\
$67.7$ & $0.7020$ & $2.1825$ & $8.55$ & $3.99$ \\
$108.4$ & $0.7360$ & $2.2665$ & $5.55$ & $2.59$ \\
\hline
$33.8$ & $0.3270$ & $2.1485$ & $14.54$ & $10.19$ \\
$45.3$ & $0.3360$ & $2.2295$ & $11.25$ & $7.88$ \\
$63.5$ & $0.3410$ & $2.2535$ & $8.10$ & $5.67$ \\
$103.3$ & $0.3340$ & $2.1745$ & $4.80$ & $3.36$ \\
\hline
\end{tabular}
\end{center}
\end{table*}

\vspace{2cm}
\begin{table}[h!]
{\bf \caption{\label{tab:OptRes}}}The results of the optimisation for fixed values of the parameter $\mu_v$ from $0.05$ to $0.50$s/m. The reduced $\chi^2$ ($\chi^2$/NDF) is also given (second column). The parameters $A$ and $B$ are correlated with 
$\mu_v$; of course, this is hardly surprising given the dependence of the tractive force on the final velocity of the vehicle (also remarked in Ref.~\cite{vt}). A dependence of both $A$ and $B$ on the weight/load of/on the vehicle is expected (the 
results of the table correspond to a $16.6$kg vehicle). The statistical uncertainties obtained from these fits are small when compared to the variation of the values with $\mu_v$, at least in the interval of interest ($0.10 \leq \mu_v \leq 0.20$s/m).
\vspace{0.6cm}
\begin{center}
\begin{tabular}{|c|c|c|c|}
\hline
$\mu_v$ (s/m) & $\chi^2$/NDF & $A$ (N$\cdot$s/m) & $B$ (N) \\
\hline
$0.05$ & $1.414124$ & $1.987763$ & $4.836199$ \\
$0.10$ & $1.323325$ & $3.845995$ & $4.838278$ \\
$0.15$ & $1.302486$ & $5.706388$ & $4.840357$ \\
$0.20$ & $1.293971$ & $7.566957$ & $4.842438$ \\
$0.25$ & $1.289486$ & $9.427339$ & $4.844528$ \\
$0.30$ & $1.286761$ & $11.288190$ & $4.846608$ \\
$0.35$ & $1.284945$ & $13.148623$ & $4.848691$ \\
$0.40$ & $1.283655$ & $15.009084$ & $4.850781$ \\
$0.45$ & $1.282694$ & $16.868739$ & $4.852889$ \\
$0.50$ & $1.281952$ & $18.729476$ & $4.854964$ \\
\hline
\end{tabular}
\end{center}
\end{table}

\clearpage
% ============= FIGURE 1

\clearpage
% ============= FIGURE 2

\clearpage
% ============= FIGURE 3

\clearpage
% ============= FIGURE 4

\clearpage
% ============= FIGURE 5

\clearpage
% ============= FIGURE 6

\clearpage
% ============= FIGURE 7

\clearpage
% ============= FIGURE 8

\clearpage
% ============= FIGURE 9

\clearpage
% ============= FIGURE 10

\clearpage
% ============= FIGURE 11

\end{document}